\documentclass[aps,showpacs,preprint,amsmath,amssymb,superscriptaddress,prl]{revtex4-1}

\usepackage{graphicx}
\usepackage{bm}
\usepackage{color} 

\begin{document}



\title{Asymmetric Pt/Co/Pt structure as a platform for extracting pure spin Hall torque}

\author{Hongyu An}
\affiliation{Department of Applied Physics and Physico-Informatics, Keio University, Yokohama 223-8522, Japan}

\author{Hiroyasu Nakayama} \affiliation{Department of Applied Physics and Physico-Informatics, Keio University, Yokohama 223-8522, Japan}

\author{Yusuke Kanno}
\affiliation{Department of Applied Physics and Physico-Informatics, Keio University, Yokohama 223-8522, Japan}

\author{Akiyo Nomura} \affiliation{Department of Applied Physics and Physico-Informatics, Keio University, Yokohama 223-8522, Japan}

\author{Satoshi Haku} \affiliation{Department of Applied Physics and Physico-Informatics, Keio University, Yokohama 223-8522, Japan}

\author{Kazuya Ando}
\email{ando@appi.keio.ac.jp}
\affiliation{Department of Applied Physics and Physico-Informatics, Keio University, Yokohama 223-8522, Japan}
\affiliation{PRESTO, Japan Science and Technology Agency, Kawaguchi, Saitama 332-0012, Japan}

\date{\today}

\begin{abstract}
We have quantitatively studied the spin-orbit torque purely generated by the spin Hall effect in a wide range of temperatures by intensionally eliminating the Rashba spin-orbit torque using Pt/Co/Pt trilayers with asymmetric thicknesses of the top and bottom Pt layers. The vanishingly small contribution from the Rashba effect has been confirmed through the vector measurements of the current-induced effective fields. In order to precisely determine the value of the spin Hall torque, the complete cancelation of the spin Hall torque has been verified by fabricating symmetric Pt/Co/Pt structure on SiO$_2$ and Gd$_{3}$Ga$_{5}$O$_{12}$ (GGG) substrates. Despite of the complete cancelation on the GGG substrate, the spin Hall torque cannot be completely canceled out even when the top and bottom Pt layers have same thicknesses on the SiO$_2$ substrate, which suggests that Pt/Co/Pt trilayers on a GGG substrate is a suitable system for precise measurements of the spin Hall torque. The result of the vector measurements on Pt/Co/Pt/GGG from 300 to 10 K shows that the spin Hall torque is almost independent of temperature, which is quantitatively reproduced under the assumption of the temperature-independent spin Hall angle of Pt.  
\end{abstract}


\maketitle

\section{I. introduction}
Magnetization switching manipulated by current in nano-scale films is promising for the future spintronic-based memory and logic devices due to its low energy consumption and high-speed response, which has attracted robust research interests in recent years~\cite{miron2011perpendicular,miron2010current,liu2012spin,liu2012current,kim2013layer,yu2014switching,fan2014magnetization,cubukcu2014spin,woo2014enhanced,zhao2015spin,garello2014ultrafast,liu2014control,yu2014magnetization,wang2014spin}. Typically, in oxide/ferromagnetic metal (FM)/non-magnetic heavy metal (NM) trilayers with perpendicular magnetic anisotropy, by injecting in-plane current into the NM layer with strong spin-orbit coupling, spin-orbit torque (SOT) is generated through the Rashba effect and spin Hall effect (SHE), then transferred into the FM layer, leading to the efficient switching of the magnetization orientation in the FM layer~\cite{akyol2015effect,yu2014current,haogiant,garello2013symmetry,avci2012magnetization,kim2014anomalous,PhysRevB.91.214434,zhang2015spin,avci2014fieldlike,hayashi2014quantitative}. In the above oxide/FM/NM material systems, two types of SOTs are responsible for the magnetization switching, the damping-like (Slonczewski) torque with the form of $\bm{\tau}_{\text{D}}\propto{\bf m}\times(\bm{\sigma}\times{\bf m})$ and the field-like torque with the form of $\bm{\tau}_{\text{F}}\propto{\bf m}\times\bm{\sigma}$, where $\bf m$ and $\bm \sigma$ are unit vectors of the magnetization in the FM layer and nonequilibrium spin polarization direction in the NM layer, respectively~\cite{garello2013symmetry,fan2013observation}. Equivalently, the two kinds of torques yield to a damping-like effective field $\bf H_\text{D}\propto\bm{\sigma}\times{\bf m}$ and a field-like effective field $\bf H_\text{F}\propto\bm{\sigma}$, respectively. The damping-like torque, which counteracts magnetization damping, can switch the magnetization directly and the field-like torque can lower the energy barrier for switching~\cite{emori2014large}. In the oxide/FM/NM trilayers, due to the structural inversion asymmetry, the Rashba effect and the SHE coexists. Both the Rashba and spin Hall effects can generate field-like $\tau_{\text{F}}$ ($\sim{H}_\text{F}$) and damping-like $\tau_{\text{D}}$ ($\sim{H}_\text{D}$) torques. 
Although this may benefit the magnetization switching, it is difficult to precisely separate and quantitatively investigate the spin-orbit physics in metallic heterostructures.

\begin{figure}[tb]
\includegraphics[scale=1]{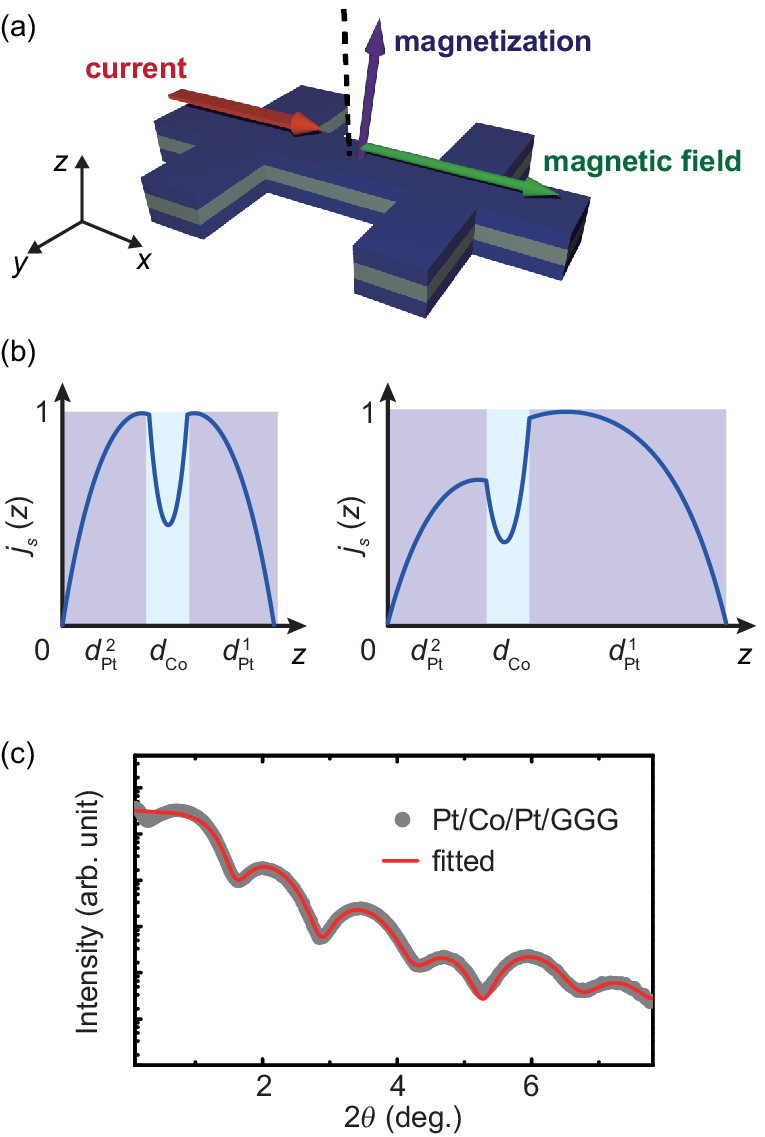}
\caption{(a) Schematic illustration of the device structure. The current is injected along $x$-axis. The external magnetic field is along $x$-axis for the magnetization switching and $H_\text{D}$ measurement, and along $y$-axis for the $H_\text{F}$ measurement. (b) The spin current $j_\text{s}(z)$ distribution diagram in Pt/Co/Pt trilayers with symmetric and asymmetric thicknesses for the two Pt layers, respectively. $d_\text{Pt}^1$ and $d_\text{Pt}^2$ represent the thicknesses of the top and bottom Pt layers. (c) The X-ray reflectivity profile for the Pt(5 nm)/Co(1.1 nm)/Pt(2 nm)/GGG trilayer. The solid circles are the experimental data and the red curve is the fitting result.}
\label{schematic} 
\end{figure}

In this paper, we fabricate Pt/Co/Pt trilayers with asymmetric thicknesses of the top and bottom Pt layers [see Fig.~\ref{schematic}(a)] to study the SOT purely generated by the SHE in a large range of temperatures. Since the Rashba SOTs, which can be induced at the top Pt/Co and bottom Co/Pt interfaces, are cancelled out due to the opposite signs of the Rashba SOTs generated at the two interfaces~\cite{yang2014platinum,haazen2013domain,sim2014asymmetry,yang2015spin}, the asymmetric Pt/Co/Pt structure allows us to quantitatively study the SOT purely generated by the SHE. In the case of symmetric thicknesses of the top and bottom Pt layers, the charge current density $j_\text{c}$ in both Pt layers will be same, leading to the complete elimination of the SOT due to the SHE. The charge current density $j_\text{c}$ 
is still same in both Pt layers even by setting different thicknesses of both Pt layers. However, the spin current density injected into the Co layer from both Pt layers will be asymmetric due to the effect of spin diffusion and relaxation as shown in Fig.~\ref{schematic}(b). Figure~\ref{schematic}(b) shows the spatial distribution of the spin current density $j_\text{s}(z)$ in the symmetric and asymmetric Pt/Co/Pt structures calculated by using the Valet-Fert equation~\cite{valet1993theory} 
\begin{equation}
\nabla^{2}\delta\mu_\text{s}=\frac{1}{\lambda^2}\delta\mu_\text{s} \label{valet}
\end{equation}
with $j_\text{s}=-({\sigma}/{2e})\nabla\delta\mu_\text{s}+\theta j_\text{c}$, where $\delta\mu_\text{s}$ is the difference in the electrochemical potential for up and down spin electrons and $\lambda$ is the spin diffusion length. $\sigma$ is the electrical conductivity and $\theta$ is the spin Hall angle. Therefore, a nonzero spin current is injected into the Co layer, generating the spin Hall torque in the asymmetric structure. Through conducting the vector measurements of the current-induced effective fields, we confirm that the Rashba SOT is negligible in the Pt/Co/Pt trilayers. 
Moreover, in order to precisely determine the value of the spin Hall torque, the complete cancelation of the spin Hall torque has been verified by fabricating symmetric Pt/Co/Pt structure on SiO$_2$ and Gd$_{3}$Ga$_{5}$O$_{12}$ (GGG) substrates. We found that despite of the complete cancelation on the GGG substrate, the spin Hall torque cannot be completely canceled out even when the top and bottom Pt layers have same thicknesses on the SiO$_2$ substrate, which suggests that Pt/Co/Pt trilayers on a GGG substrate is a suitable system for precise measurements of the SOT purely generated by the SHE. The dependence of the spin Hall torque with temperatures has also been clarified.

\section{II. experimental methods}
For the sample fabrication, Pt/Co/Pt films were deposited on a thermally oxidized Si (SiO$_2$) substrate and a Gd$_{3}$Ga$_{5}$O$_{12}$ (GGG) (111) substrate at room temperature by rf magnetron sputtering. Before the deposition, the films were patterned into 10 $\mu$m$\times$100 $\mu$m Hall bar shape by standard photolithography, and lift off technique was used to take off the rest part of the films after deposition. To quantitatively evaluate the roughness at the top Pt/Co and bottom Co/Pt interfaces, we measured X-ray reflectivity (XRR) with a Bruker D8 Discover diffractometer by applying Cu K$_\alpha$ radiation for a Pt(5 nm)/Co(1.1 nm)/Pt(2 nm)/GGG trilayer. As shown in Fig.~\ref{schematic}(c), the high-order peaks of the Kiessig fringes can be clearly detected, which indicates smooth interfaces and surface conditions. The measured spectrum was fitted using the Bruker LEPTOS software with the Levenberg-Marquardt method. With the best fitting result, the roughness at the Pt/Co and Co/Pt interfaces was obtained as 0.8 $\rm{\AA}$ and 0.9 $\rm{\AA}$, respectively. This result indicates little difference of the roughness between the Pt/Co and Co/Pt interfaces. The magnetization and transport measurement were conducted using Physical Property Measurement System (PPMS, Quantum Design) in temperature range of 10 to 300 K. The vector measurements were taken by using two lock-in amplifiers with a current frequency of 507.32 Hz.

\section{III. results and discussion}
\begin{figure}[tb]
\includegraphics[scale=1]{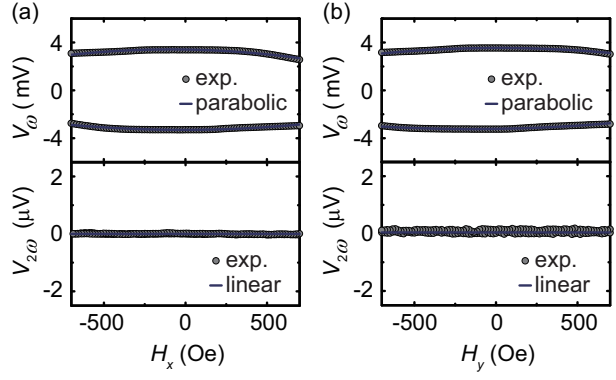}
\caption{The first $V_\omega$ and second $V_{2 \omega}$ harmonic Hall voltage signals as a function of the in-plane magnetic field applied along (a) the $x$-axis and (b) the $y$-axis for the symmetric Pt(3.5 nm)/Co(1.1 nm)/Pt(3.5 nm)/GGG trilayer. The applied current density was $J=0.42\times 10^6$ A/cm$^2$. The solid circles are the experimental data. The solid lines are the fitting results using parabolic and linear functions. }
\label{harmonic} 
\end{figure}

In order to confirm that the SOTs due to the Rashba effect is eliminated in our designed devices, vector measurements of the current-induced effective fields were conducted for the symmetric Pt(3.5 nm)/Co(1.1 nm)/Pt(3.5 nm) trilayer on the GGG substrate. By applying an a.c. current, the first $V_\omega$ and second $V_{2\omega}$ harmonic Hall voltage signals were simultaneously measured by using two lock-in amplifiers as shown in Figs.~\ref{harmonic}(a) and \ref{harmonic}(b). The Hall voltage typically contains contributions from the anomalous Hall effect (AHE) and the planar Hall effect (PHE). From the magnetic field dependence of $V_\omega$ and $V_{2\omega}$, the effective fields associated with the field-like torque $H_\text{F}$ and damping-like torque $H_\text{D}$ were calculated using the following equation in which the PHE contribution is included~\cite{hayashi2014quantitative}:
\begin{equation}
H_\text{F(D)}=\frac{H'_\text{F(D)}\pm2\xi H'_\text{D(F)}}{1-4\xi^2},\label{harmoniceq}
\end{equation}
where $\xi=\Delta R_\text{PHE}/\Delta R_\text{AHE}$ and 
\begin{equation}
H'_\text{F(D)}=-2\left(\frac{\partial V_{2\omega}}{\partial H_{y(x)}}\right)/\left(\frac{\partial^2 V_{\omega}}{\partial H^2_{y(x)}}\right). \label{HFD}
\end{equation}
Here, $H_{y}$ and $H_{x}$ are the applied external magnetic field in transverse (align with the $y$-axis) and longitudinal (align with the $x$-axis) direction. The Hall resistance due to the PHE, $\Delta R_\text{PHE}$, and that due to the AHE, $\Delta R_\text{AHE}$, were measured for the Pt/Co/Pt devices. The determined $\xi=\Delta R_\text{PHE}/\Delta R_\text{AHE}$ are $\xi=0.125$ for the Pt/Co/Pt/GGG trilayer and $\xi=0.121$ for the Pt/Co/Pt/SiO$_2$ trilayer. Thus, the denominator of Eq.~(\ref{harmoniceq}) for the Pt/Co/Pt trilayers used in the present study is $1-4\xi^2\simeq 0.94$ and $H_\text{F(D)}$ can be approximated to be 
\begin{equation}
H_\text{F(D)}\simeq{H'_\text{F(D)}\pm2\xi H'_\text{D(F)}}.\label{HFD2}
\end{equation}
As shown in Figs.~\ref{harmonic}(a) and \ref{harmonic}(b), the first harmonic voltage $V_{\omega}$ signal is clearly observed for the symmetric Pt/Co/Pt/GGG trilayer both when the external magnetic field is applied along the $x$ and $y$ axes. However, we observed no obvious second harmonic voltage $V_{2\omega}$: $\partial V_{2\omega}/\partial  H_{y(x)}=0$. This demonstrates that both the damping-like $H_\text{D}$ and field-like $H_\text{F}$ effective fields are vanishingly small in the present system [see Eqs.~(\ref{HFD}) and (\ref{HFD2})].

The negligible damping-like and field-like SOTs in the Pt/Co/Pt/GGG trilayer indicates that the Rashba SOTs, which can be induced at the top Pt/Co and bottom Co/Pt interfaces, are eliminated because of the symmetric structure. In the symmetric Pt/Co/Pt trilayer, the damping-like and field-like torques due to the SHE is expected to disappear because the net spin current injected into the Co layer is zero as shown in Fig.~\ref{schematic}(b). The damping-like and field-like torques due to the Rashba effect are also expected to be cancelled out due to the opposite signs of the Rashba SOTs generated at the top Pt/Co and bottom Co/Pt interfaces. However, the cancelation of the Rashba SOTs can be incomplete when the two interfaces are not identical. Thus, the negligible SOTs in the symmetric Pt/Co/Pt trilayer can also be explained as that the non-negligible Rashba SOTs due to the incomplete cancelation is canceled by nonzero SOTs due to the SHE because of possible asymmetry in the top and bottom Pt layers. However, this possibility is unlikely to be origin of the negligible SOTs in the symmetric Pt/Co/Pt trilayer, since this situation has to satisfy two strict requirements: firstly, the damping-like torques generated by the Rashba and SHE have the same magnitude and opposite sign; secondly, the field-like torques generated by the Rashba and SHE also have the same magnitude and opposite sign. Therefore, the negligible damping-like and field-like torques in the symmetric Pt/Co/Pt structure suggests that both the SOTs generated by the Rashba and SHE are vanishingly small; although the Rashba SOTs can be induced at the Pt/Co interface, they are largely eliminated in the present system because of the nearly identical Pt/Co and Co/Pt interfaces, which is supported by the XRR data shown in Fig.~\ref{schematic}(c). Although the field-like torque has been observed in metallic heterostructures previously, most of the previous studies used asymmetric devise structures, such as Pt/Co/MgO, Ta/CoFeB/MgO, Pt/Co/AlO$_x$, and Pt/Co/Ta, which makes it difficult to separate the Rashba and spin Hall torques in these devices. In contrast to these studies, in the present study, we use the symmetric Pt/Co/Pt structure to minimize SOTs arising from the inversion asymmetry, which would provide a simpler platform for studying the SOTs arising from the SHE.

\begin{figure}[tb]
\includegraphics[scale=1]{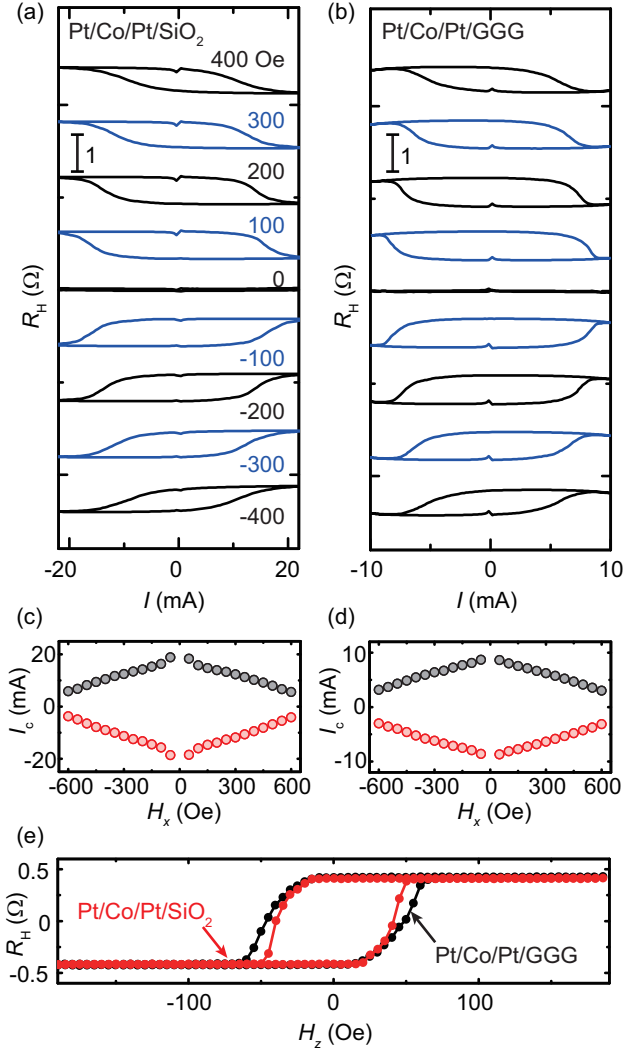}
\caption{(a) and (b) Current-induced magnetization switching curves for the asymmetric Pt/Co/Pt/SiO$_2$ and Pt/Co/Pt/GGG with different in-plane external magnetic field $H_x$, respectively. (c) and (d) Switching phase diagrams for Pt/Co/Pt/SiO$_2$ and Pt/Co/Pt/GGG, respectively. (e) Anomalous Hall resistance $R_\text{H}$ dependence on the perpendicular magnetic field.}
\label{switching} 
\end{figure}

Next, to induce nonzero SOTs generated by the SHE, we use asymmetric Pt/Co/Pt structures. Figures~\ref{switching}(a) and \ref{switching}(b) show the current-induced magnetization switching for the asymmetric Pt(5.0 nm)/Co(1.1 nm)/Pt(2.0 nm)/SiO$_2$ and Pt(5.0 nm)/Co(1.1 nm)/Pt(2.0 nm)/GGG devices measured using the anomalous Hall effect (AHE); the anomalous Hall resistance $R_\text{H}$ was measured as a function of an in-plane dc current with applying an in-plane external magnetic field along $x$-axis [see Fig.~\ref{schematic}(a)]. In order to avoid thermal effects generated by the current, the experiment was conducted at 250 K. As shown in Figs.~\ref{switching}(a) and \ref{switching}(b), by reversing the direction of the external in-plane magnetic field, the polarity of the magnetization switching is also reversed, which is consistent with the SHE-induced magnetization switching~\cite{liu2012current}. The corresponding switching phase diagrams (SPDs) are plotted in Figs.~\ref{switching}(c) and \ref{switching}(d). Both SPDs show almost symmetric shapes, indicating the uniform reversal process~\cite{liu2012current}. By increasing the value of the in-plane magnetic field, the critical current decreases accordingly. With an in-plane magnetic field $H_{x}$=100 Oe, the critical switching current $I_\text{c}$ is approximately 16 mA for the Pt/Co/Pt/SiO$_2$ device, which yields to a current density of 1.97$\times10^7$ A/cm$^2$. For the Pt/Co/Pt/GGG device, $I_\text{c}$ is only 8 mA in the same condition of the in-plane magnetic field. By comparing the $R_\text{H}$ dependence on the perpendicular magnetic field of the two samples [see Fig.~\ref{switching}(e)], it is found that the Pt/Co/Pt/GGG device has a smaller slope during the magnetization switching, indicating a weaker perpendicular anisotropy field than the Pt/Co/Pt/SiO$_2$ device. The weaker perpendicular anisotropy field results in a reduction of the critical switching current. 
By increasing $H_x$ to above 400 Oe, the magnetization switching becomes incomplete due to the too large tilting in the in-plane direction.

\begin{figure}[tb]
\includegraphics[scale=1]{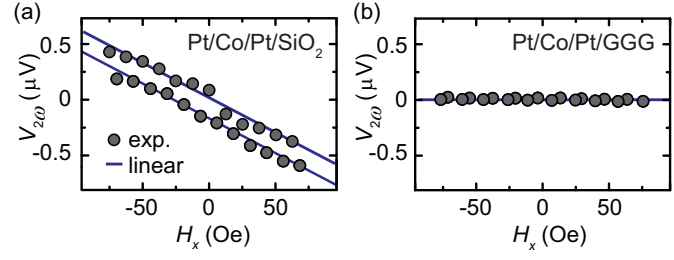}
\caption{The in-plane magnetic field dependence of $V_{2\omega}$ for the symmetric Pt/Co/Pt structure on the (a) SiO$_2$ and (b) GGG substrates. The in-plane magnetic field was applied along the $x$-axis.}
\label{SiO2GGG} 
\end{figure}

The above experimental results demonstrate that the Rashba effect is eliminated and nonzero spin Hall torque is responsible for the manipulation of the magnetization in the asymmetric Pt/Co/Pt structures. Here, the top and bottom Pt layers generate the damping-like torque in opposite directions (i.e., defined as $\tau_\text{t}$ and $\tau_\text{b}$), and the net effective field~\cite{liu2012current}
\begin{equation}
H_\text{D}=H_\text{t}-H_\text{b}=\frac{\hbar}{2 e M_\text{s} d_\text{F}} (j_\text{t}-j_\text{b}) \label{torque}
\end{equation}
after the cancellation from both sides acts on the magnetization in the Co layer, where $M_\text{s}$ and $d_\text{F}$ are the saturation magnetization and thickness of the Co layer, respectively. $H_\text{t}$ and $H_\text{b}$ are the effective fields associated with $\tau_\text{t}$ and $\tau_\text{b}$, respectively. $j_\text{t}$ and $j_\text{b}$ are the spin current density $j_\text{s}(z)$ injected into the Co layer from the top and bottom Pt layers. For the Pt/Co/Pt trilayers, $H_\text{t}$ and $H_\text{b}$ are purely generated by the SHE because of the vanishingly small Rashba SOT. Thus, by measuring $H_\text{D}$ and by solving Eq.~(\ref{valet}), the SHE in the Pt layers can be quantified. However, we found that the Pt/Co/Pt trilayer deposited on the SiO$_2$ substrate is not a suitable system for precise measurements of $H_\text{D}$. Figure~\ref{SiO2GGG}(a) shows the magnetic field dependence of the second harmonic Hall voltage $V_{2\omega}$ for the symmetric Pt(3.5 nm)/Co(1.1 nm)/Pt(3.5 nm)/SiO$_2$ trilayer, where the in-plane magnetic field was applied along the $x$-axis. For comparison, we show the $H_x$ dependence of $V_{2\omega}$ for the symmetric Pt(3.5 nm)/Co(1.1 nm)/Pt(3.5 nm)/GGG trilayer in Fig.~\ref{SiO2GGG}(b). Figure~\ref{SiO2GGG}(a) shows that $V_{2\omega}$ decreases clearly with $H_x$ for the symmetric Pt/Co/Pt/SiO$_2$ trilayer. This indicates that the cancellation of the spin Hall torque from the top and bottom Pt layers is not complete in the Pt/Co/Pt/SiO$_2$ device. By using the Eq.~(\ref{harmoniceq}), $H_\text{D}$ is calculated to be $1.09\times 10^{-6}$ Oe/Acm$^{-2}$ for the symmetric Pt/Co/Pt/SiO$_2$ device. The sign of the $H_\text{D}$ suggests that the magnitude of $\tau_\text{t}$ is larger than that of $\tau_\text{b}$ in spite of the same thicknesses of the top and bottom Pt layers on SiO$_2$ substrate. Although the reason is not clear currently, this may be due to the influence of the SiO$_2$ substrate. We fabricated Pt single layer films (3.5 nm) on SiO$_2$ and GGG substrates simultaneously, and measured the resistivity by using four probes method, respectively. It is found that the resistivity of Pt on the SiO$_2$ substrate (63.20 $\mu\Omega$cm) is larger than which on the GGG substrate (50.19 $\mu\Omega$cm). According to the previous study, a thin Pt film deposited on a SiO$_2$ substrate is lack of strong adhesion between the Pt atoms and the SiO$_2$ surface, which leads to a increase of the Pt resistivity~\cite{PhysRevLett.116.126601}. This result is consistent with the vector measurement on the symmetric Pt/Co/Pt trilayers. Since in the symmetric Pt/Co/Pt/SiO$_2$ device, if the bottom Pt layer has a larger resistivity than the top Pt layer, the magnitude of $\tau_\text{t}$ will be larger than $\tau_\text{b}$ because of the asymmetry of the charge current density. SiO$_2$ substrates have been commonly used for fabricating spintronic devices and the magnitude of the SOT in these devices has been under debate recently. Our results reveal that carefully controlled experiments in different substrates are necessary to fully understand the spin-orbit physics in metallic heterostructures.





\begin{figure}[tb]
\includegraphics[scale=1]{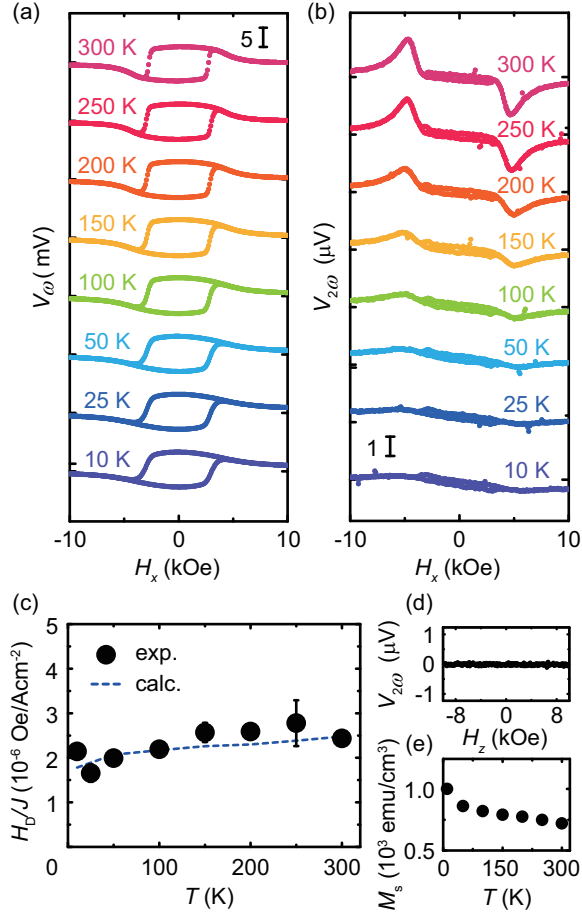}
\caption{The temperature $T$ dependence of the (a) $V_\omega$ and (b) $V_{2 \omega}$ for the asymmetric Pt/Co/Pt/GGG device measured with the in-plane magnetic field along the $x$-axis. (c) The $T$ dependence $H_\text{D}/J$, where $H_\text{D}$ is the damping-like effective field. (d) $H_z$ dependence of $V_{2\omega}$ for the asymmetric Pt/Co/Pt/GGG trilayer. (e) The $T$ dependence of the saturation magnetization measured with a vibrating sample magnetometer.}
\label{temp} 
\end{figure}

The complete cancellation of the SOTs from the top and bottom Pt layers, shown in Fig.~\ref{harmonic}, suggests that a Pt/Co/Pt trilayers on a GGG substrate is a suitable system for precise measurements of the spin Hall torque; the damping-like torque purely generated by the SHE can be quantified under the vanishingly small Rashba SOT by fabricating the trilayer with asymmetric thicknesses of the two Pt layers on a GGG substrate. In Figs.~\ref{temp}(a) and \ref{temp}(b), we show the magnetic field dependence of the first $V_\omega$ and second $V_{2\omega}$ harmonic Hall voltage signals for the asymmetric Pt(5.0 nm)/Co(1.1 nm)/Pt(2.5 nm)/GGG device at various temperatures $T$, where the in-plane magnetic field was applied along the $x$-axis. Using the result shown in Figs.~\ref{temp}(a) and \ref{temp}(b) combined with Eq.~(\ref{HFD2}), the temperature dependence of the damping-like effective field $H_\text{D}$ is obtained as shown in Fig.~\ref{temp}(c). The effective field is almost constant $H_\text{D}/J\sim 2.5\times 10^{-6}$ Oe/Acm$^{-2}$ and it decreases only slightly with decreasing the temperature $T$. Here, in the Hall measurements, the anomalous Nernst effect (ANE) can also be induced by current-induced inhomogeneous sample heating, which gives rise to additional contribution to the second harmonic voltage $V_{2\omega}$. The ANE contribution to $V_{2\omega}$ can be determined by measuring the harmonic voltage as a function of the magnetic field along the $z$ direction, $H_z$~\cite{garello2013symmetry}. However, we observed that $V_{2\omega}$ is negligible for the asymmetric Pt/Co/Pt structure as shown in Fig.~\ref{temp}(d), and thus we neglected the ANE contribution to $V_{2\omega}$ in the present system.

To quantitatively discuss the temperature dependence of the damping-like torque shown in Fig.~\ref{temp}(c), we have solved the spin diffusion equation, Eq.~(\ref{valet}). By taking into account the SHE in the top and bottom Pt layers with the boundary conditions in Pt/Co/Pt trilayers, i.e. $j_\text{s}=0$ at the surfaces of the Pt layers and continuity of $j_\text{s}$ and $\delta\mu_\text{s}$ at the Pt/Co and Co/Pt interfaces, the net spin current density $j_\text{s}={j_\text{t}} - {j_\text{b}}$ injected into the Co layer is obtained as

\begin{eqnarray}
\scalebox{0.75}{$\displaystyle
{j_\text{s}} = \frac{{4\theta \lambda_\text{N} \sigma_{\rm{F}}\sinh \left( {\frac{d_\text{F}}{{{\rm{2}}\lambda _\text{F}}}} \right)\sinh \left( {\frac{{{{d_\text{N}^1 - d_\text{N}^2}}}}{{{\rm{2}}\lambda_\text{N} }}} \right)\left[ {{\rm{2}}\lambda _\text{F}\sigma _\text{N}\sinh \left( {\frac{d_\text{F}}{{{\rm{2}}\lambda _\text{F}}}} \right)\sinh \left( {\frac{{{{d_\text{N}^1}}}}{{{\rm{2}}\lambda_\text{N} }}} \right)\sinh \left( {\frac{{{{d_\text{N}^2}}}}{{{\rm{2}}\lambda_\text{N} }}} \right){\rm{ + }}\lambda_\text{N} \sigma_{\rm{F}}\cosh \left( {\frac{d_\text{F}}{{{\rm{2}}\lambda _\text{F}}}} \right)\sinh \left( {\frac{{{{d_\text{N}^1 + d_\text{N}^2}}}}{{{\rm{2}}\lambda_\text{N} }}} \right)} \right]}}{{\sinh \left( {\frac{d_\text{F}}{{\lambda _\text{F}}}} \right)\left[ {{\lambda_\text{N} ^2}\sigma_\text{F} {^{\rm{2}}}\cosh \left( {\frac{{{{d_\text{N}^1}}}}{\lambda_\text{N} }} \right)\cosh \left( {\frac{{{{d_\text{N}^2}}}}{\lambda_\text{N} }} \right){\rm{ + }}\lambda_\text{F}^{\rm{2}}{\sigma_\text{N} ^{\rm{2}}}\sinh \left( {\frac{{{{d_\text{N}^1}}}}{\lambda_\text{N} }} \right)\sinh \left( {\frac{{{{d_\text{N}^2}}}}{\lambda_\text{N} }} \right)} \right] + \lambda_\text{N} \lambda _\text{F}\sigma_\text{N} \sigma_{\rm{F}}\cosh \left( {\frac{d_\text{F}}{{\lambda _\text{F}}}} \right)\sinh \left( {\frac{{{{d_\text{N}^1 + d_\text{N}^2}}}}{\lambda_\text{N} }} \right)}}j_\text{c}^\text{N},
\label{spincurrent}
$}
\end{eqnarray}
where $\theta $ is the spin Hall angle of Pt. $\lambda_\text{N(F)}$ and $\sigma_\text{N(F)}$ are the spin diffusion length and electrical conductivity of Pt (Co). $d_\text{F}$ and $d_\text{N}^{1(2)}$ are the thickness of the Co layer and top (bottom) Pt layer, respectively. $j_\text{c}^\text{N}$ is the applied charge current density in the Pt layer. The experimentally measured $H_\text{D}/J\sim 2.5\times 10^{-6}$ Oe/Acm$^{-2}$ is quantitatively reproduced using Eqs.~(\ref{torque}) and (\ref{spincurrent}) with $\theta=0.24$, $\lambda_\text{N} = 1.5$ nm, $\lambda_\text{F}=0.4$ nm, $\sigma_\text{N}=1.4\times 10^6$ $\Omega^{-1}$m$^{-1}$, $\sigma_\text{F}=3.3\times 10^6$ $\Omega^{-1}$m$^{-1}$~\cite{PhysRevB.87.224401, ghosh2012penetration}. Here, $M_s = 720$ emu/cm$^3$, measured with a vibrating sample magnetometer, was used for the calculation. The slight decrease of the damping-like torque, purely generated by the SHE, by decreasing $T$ is mainly due to the temperature dependence of the saturation magnetization $M_s$ in the Co layer. As shown in Fig.~\ref{temp}(c), by assuming negligible temperature dependence of the other parameters, which is practical~\cite{meyer2014temperature}, the $T$ dependence of $H_\text{D}/J$ is well reproduced with Eqs.~(\ref{torque}) and (\ref{spincurrent}) using the measured $M_{s}$ at each temperature [Fig.~\ref{temp}(e)]. These results demonstrate the temperature-independent spin Hall angle of Pt. It should be noticed that the above calculated spin Hall angle of Pt here sets a lower bound, since the spin transparency and the spin memory loss (SML) at the Pt/Co interface have been neglected. Therefore, the actual internal spin Hall angle of Pt would be even larger. According to the previous work, the spin transparency at a Pt/Co interface is evaluated as $T = 0.65$~\cite{zhang2015role}. By taking into account this value, the spin Hall angle of Pt in our devices is calculated to be around 0.36. This value is much larger than the previous reports on the spin Hall angle of Pt without considering the SML, while consistent with Pai $et$ $al.$'s recent study, which reported that the spin Hall angle of Pt could be as high as 0.30 or even greater~\cite{pai2015dependence}. This suggets that by engineering the NM/FM interface, spintronics devices with high conversion effeciency might be realized.

\section{IV. conclusions}
In summary, we have demonstrated a quantitative study of the pure spin Hall torque in a large range of temperatures from 300 to 10 K. By intentionally fabricating Pt/Co/Pt trilayers with asymmetric thicknesses of the top and bottom Pt layers, the Rashba spin-orbit torques are largely eliminated. In order to precisely determine the value of the spin Hall torque, the complete cancelation of spin Hall torque has been verified by fabricating symmetric Pt/Co/Pt structure on SiO$_2$ and GGG substrates. It is found that despite of the complete cancelation on the GGG substrate, the spin Hall torque cannot be completely canceled out when the top and bottom Pt layers have same thicknesses on the SiO$_2$ substrate. This suggests that Pt/Co/Pt trilayers on a GGG substrate is a suitable system for precise measurements of the spin Hall torque. The result of the vector measurements for Pt/Co/Pt/GGG shows that the spin Hall angle of Pt is independent with the temperatures.

\section{acknowledgments}

\begin{acknowledgments}
This work was supported by PRESTO-JST ``Innovative nano-electronics through interdisciplinary collaboration among material, device and system layers," JSPS KAKENHI Grant Numbers 26220604, 26103004, 26600078, 26790037, the Mitsubishi Foundation, the Asahi Glass Foundation, the Mizuho Foundation for the Promotion of Sciences, and the Casio Science Promotion Foundation. H. A. and H. N. contributed equally to this work.
\end{acknowledgments}

\clearpage


%

\clearpage

\clearpage

\clearpage

\clearpage

\clearpage

\end{document}